\documentclass[aps,prl,reprint,amsmath,amssymb, superscriptaddress]{revtex4-2} %
\usepackage{amsmath} %
\usepackage{bm}  %
\usepackage{graphicx}
\usepackage{physics}
\usepackage{color}
\usepackage{braket}
\usepackage{soul}

\usepackage[colorlinks=true, citecolor=blue, linkcolor=blue, urlcolor=blue]{hyperref}

\newcommand{\kk}{\bm{k}}
\renewcommand{\qq}{\bm{q}}

\begin{document}

\title{Predicting Phonon-Induced Spin Decoherence from First-Principles: \\
Colossal Spin Renormalization in Condensed Matter}
\author{Jinsoo Park}
\affiliation{Department of Applied Physics and Materials Science, California Institute of Technology, Pasadena, CA 91125, USA.}
\author{Jin-Jian Zhou}
\affiliation{School of Physics, Beijing Institute of Technology, Beijing 100081, China.}
\author{Yao Luo}
\affiliation{Department of Applied Physics and Materials Science, California Institute of Technology, Pasadena, CA 91125, USA.}
\author{Marco Bernardi}
\email[Corresponding author: ]{bmarco@caltech.edu}
\affiliation{Department of Applied Physics and Materials Science, California Institute of Technology, Pasadena, CA 91125, USA.}
\begin{abstract}
Developing a microscopic understanding of spin decoherence is essential to advancing quantum technologies. %
Electron spin decoherence due to atomic vibrations (phonons) plays a special role as it sets an intrinsic limit to the performance of spin-based quantum devices. 
Two main sources of phonon-induced spin decoherence $–$ the Elliott-Yafet (EY) and Dyakonov-Perel (DP) mechanisms $–$ have distinct physical origins and theoretical treatments. %
Here we show calculations that unify their modeling  
and enable accurate predictions of spin relaxation and precession in semiconductors. %
We compute the phonon-dressed vertex of the spin-spin correlation function with a treatment analogous to the calculation of the anomalous electron magnetic moment in QED. %
We find that the vertex correction provides a giant renormalization of the electron spin dynamics in solids, greater by many orders of magnitude than the corresponding correction from photons in vacuum. 
Our work demonstrates a general approach for quantitative analysis of spin decoherence in materials, advancing the quest for spin-based quantum technologies.%
\end{abstract} %
\maketitle
\begin{figure}[t!]
\includegraphics[width=0.95\columnwidth]{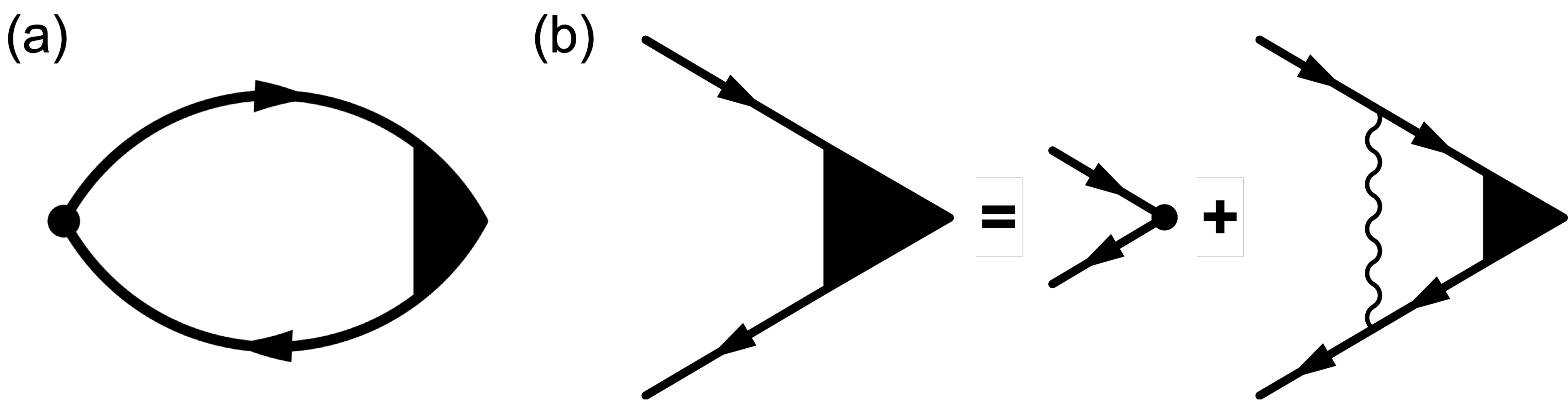}
\caption{Feynman diagrams for spin decoherence. (a) Bubble diagram for the spin-spin correlation function including the vertex correction. (b) Bethe-Salpeter equation for the phonon-dressed spin vertex in the ladder approximation.
}\label{fig:diagram}
\end{figure}
\begin{figure*}[t!]
\includegraphics[width=1.0\textwidth]{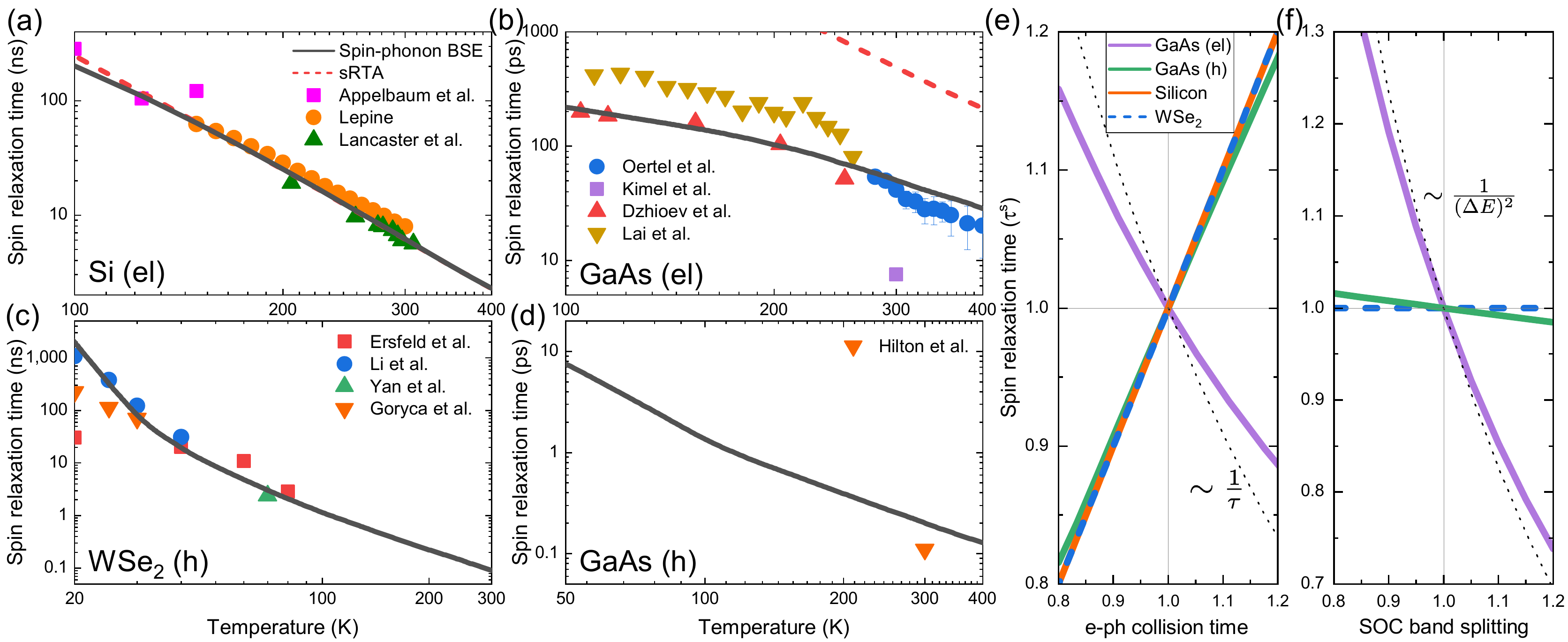} %
\caption{Spin relaxation times. (a)-(d) Computed spin relaxation times as a function of temperature, for (a) electrons in Si, (b) electrons in GaAs, (c) holes in monolayer $\text{WSe}_2$, and (d) holes in GaAs. Results obtained from the spin-phonon BSE (black solid line) are compared with sRTA calculations (red dashed line). 
Experimental results from Refs.~\cite{appelbaumElectronic2007,lepineSpin1970,lancasterSpinlattice1964,oertelHigh2008,kimelRoomtemperature2001,dzhioevSuppression2004,laiTemperature2006,ersfeldUnveiling2020,liValley2021,yanLong2017,gorycaDetection2019} are shown for comparison. 
(e)-(f)~The SRTs at room temperature for these four cases are recomputed by artificially varying (e) the $e$-ph collision time and
(f) the SOC band splitting entering the spin-phonon BSE.
In all cases, the axes are referenced to the real system values. The conventional DP spin relaxation trend (black dotted line) is also shown for comparison. %
}\label{fig:srt_temp}
\end{figure*}

Spin decoherence from phonons is a pressing question in quantum technology $-$ it governs spin transport~\cite{balasubramanianNanoscale2008,jansenSilicon2012,zuticSpintronics2004,hanGraphene2014,appelbaumElectronic2007,zeleznySpin2018} and limits the manipulation of quantum information~\cite{veldhorstTwoqubit2015,miCoherent2018,whiteleySpin2019,wolfowiczQuantum2021,noiriFast2022,bourassaEntanglement2020,petitUniversal2020} and the realization of reliable quantum devices~\cite{degenQuantum2017,vasyukovScanning2013,barrySensitivity2020}.
Previous work has identified two key sources of phonon-induced spin decoherence in the presence of spin-orbit coupling (SOC) $-$ the Elliott-Yafet (EY) mechanism~\cite{elliottTheory1954,yafetFactors1963}, whereby electron-phonon ($e$-ph) collisions change the spin direction, and the Dyakonov-Perel (DP) mechanism~\cite{dyakonov1972spin} originating from spin precession between $e$-ph collisions. 
Historically, these two mechanisms have been described with distinct theoretical models~\cite{elliottTheory1954,yafetFactors1963,dyakonov1972spin,baralReexamination2016,mowerDyakonovPerel2011}, but significant efforts have been made to unify them, for example using real-time evolution of spin ensembles~\cite{wuSpin2010,xuInitio2021,shenHole2010} or analyzing quasiparticle broadening in model systems~\cite{simonGeneralized2008,borossUnified2013,szolnokiIntuitive2017}.
\\
\indent
However, formulating a theory that encompasses both the EY and DP mechanisms, and developing corresponding quantitative calculations of spin decoherence in real materials, are still outstanding challenges.
Many-body approaches combined with density functional theory (DFT) and related first-principles calculations are particularly promising to tackle this problem. 
These \textit{ab initio} methods have become a gold standard for calculations of $e$-ph interactions and transport phenomena in solids~\cite{bernardiFirstprinciples2016,zhouPerturbo2021,sohierPhononlimited2014,liElectrical2015,zhouInitio2016,poncePredictive2018,zhouElectronPhonon2018,jhalaniPiezoelectric2020,zhouInitio2021}. %
Recent work has extended this framework to compute spin-flip processes due to $e$-ph interactions, leading to predictions of EY spin decoherence within the spin relaxation time approximation (sRTA)~\cite{parkSpinphonon2020}. 
It is widely accepted that the sRTA neglects spin precession, and thus a different formalism is needed to capture the DP mechanism~\cite{baralReexamination2016,shenHole2010}. 
\\
\indent
Inspired by the work of Kim et al.~\cite{kimVertex2019}, which rigorously proved that the Boltzmann equation is equivalent to the ladder vertex correction to the conductivity, we ask if a similar many-body approach can be used to study spin dynamics. %
The development of this framework, and of corresponding first-principles calculations, would provide a viable tool to study phonon-induced spin decoherence, mimicking the progress of first-principles studies of charge transport~\cite{bernardiFirstprinciples2016,zhouPerturbo2021,sohierPhononlimited2014,liElectrical2015,zhouInitio2016,poncePredictive2018,zhouElectronPhonon2018,jhalaniPiezoelectric2020,zhouInitio2021}. 
In turn, accurate predictions of spin decoherence would advance both condensed matter theory and spin-based quantum technology. %
\\
\indent
Here we present a many-body theory of spin relaxation and develop precise \textit{ab initio} calculations of phonon-induced spin decoherence in semiconductors. 
Our approach calculates the $e$-ph vertex corrections to the spin susceptibility, with an accurate account of electronic and vibrational states, SOC, and $e$-ph interactions. 
We compute the spin relaxation times (SRTs) of electron and hole carriers in Si and GaAs $-$ two key candidates for spin-based quantum computing $-$ and in monolayer WSe$_2$, a 2D semiconductor with strong SOC.
Our predicted SRTs are in excellent agreement with experiments over a wide temperature range. We demonstrate that our formalism can calculate both spin relaxation and spin precession, and capture EY and DP decoherence on equal footing; we contrast these results with the sRTA, which lacks DP decoherence and gives unphysical SRTs near the band gap. 
Our analysis shows that the $e$-ph interactions lead to a colossal renormalization of the electron spin dynamics in solids, significantly modifying the SRTs and spin precession rates (SPRs).
The theory and computational method developed in this work pave the way for a deeper understanding of electron spin decoherence, with broad implications for quantum materials and devices.
\\ 
\indent
To describe phonon-induced spin decoherence, we consider the Kubo formula for the spin-spin correlation function~\cite{mahanManyParticle2000}, and include the ladder vertex correction~\cite{kimVertex2019} from $e$-ph interactions (see Fig.~\ref{fig:diagram}(a)).
We derive a Bethe-Salpeter equation for the phonon-dressed spin vertex (in short, spin-phonon BSE), as discussed in the Supplemental Material~\cite{supp_mat} and in the companion paper~\cite{companion}. Our spin-phonon BSE is shown diagrammatically in Fig.~\ref{fig:diagram}(b), and can be written as:
\begin{equation} \label{eq:bse}
\begin{split}
    \bm{s}\bm{\Lambda}_{\kk}(\varepsilon)\!=\!\bm{s}_{\kk}+\frac{1}{V}\!\sum_{\nu\qq\pm} 
    &\textbf{g}_{\nu\kk\qq}^\dagger 
    \!\left[\textbf{G}^A\bm{s}\bm{\Lambda} 
     \textbf{G}^R \right]_{\!\!\begin{smallmatrix}\kk+\qq,~~ \\ \varepsilon\pm\omega_{\nu\qq}\end{smallmatrix}} 
     \!\!\!\textbf{g}_{\nu\kk\qq}\, F_{\pm}(T) \\
\end{split}
\vspace{-15pt}
\end{equation} 
where all bolded quantities are matrices in Bloch basis. Above, $\bm{s}\bm{\Lambda}_{\kk}(\varepsilon) = \bm{s}_{nn'\kk} \bm{\Lambda}_{nn'\kk}(\varepsilon)$ is the phonon-dressed spin vertex,  $\Lambda^\alpha_{nn'\kk}(\varepsilon)$ is the vertex correction at energy $\varepsilon$ for the Cartesian direction $\alpha$, and $\bm{s}_{nn'\kk} = \braket{n'\kk | \tfrac{\hbar}{2}\hat{\sigma} | n\kk}$ is the bare spin vertex;
$\textbf{G}^{R/A}$ are the retarded/advanced interacting Green's functions~\cite{mahanManyParticle2000}, $V$ is the system volume, $F_{\pm}(T)$ is a thermal occupation factor at temperature $T$, and $\left[\mathbf{g}_{\nu\kk\qq}\right]_{nm}=g_{nm\nu}(\kk,\qq)$ are $e$-ph matrix elements~\cite{zhouPerturbo2021}.
\\
\indent
The vertex correction $\Lambda$ governs the spin dynamics by renormalizing the microscopic SRTs and SPRs~\cite{companion,supp_mat}. 
The macroscopic SRTs are obtained as the thermal average
\begin{equation}\label{eq:srt}
    \tau^{(s)}_{\alpha\beta}=\frac{\sum_{n\kk} s^\alpha_{nn\kk}s^\beta_{nn\kk}\tau^\text{e-ph}_{n\kk}\Lambda^{\beta}_{nn\kk}(\varepsilon_{n\kk}) (-\frac{df_{n\kk}}{d\varepsilon})}{\sum_{n\kk}   s^\alpha_{nn\kk}s^\beta_{nn\kk}
    \,\big(\!-\frac{df_{n\kk}}{d\varepsilon} \big)},
\end{equation}
where $\tau^\text{e-ph}_{n\kk}$ are $e$-ph collision times~\cite{mahanManyParticle2000,bernardiFirstprinciples2016}. For $\alpha=\beta$ along the external magnetic field, Eq.~(\ref{eq:srt}) gives the longitudinal SRT, usually called $T_1$, along the direction $\alpha$, while for %
a perpendicular magnetic field one obtains the transverse SRT, $T_2$ (not computed here)~\cite{burkovSpin2004}.
The renormalized microscopic SRTs ($\tau_{nn'\kk}^\alpha$) and SPRs ($\omega_{nn'\kk}^\alpha$), which are matrices in Bloch basis, are computed from the vertex correction using 
\begin{equation}\label{eq:dressed_time}
    \frac{1}{\frac{1}{\tau_{nn'\kk}^\alpha(\varepsilon)}+i{\omega}_{nn'\kk}^\alpha(\varepsilon)}\equiv\frac{\Lambda^{\alpha}_{nn'\kk}(\varepsilon)}{i(\Sigma^R_{n\kk}-\Sigma^A_{n'\kk})+i(\varepsilon_{n\kk}-\varepsilon_{n'\kk})},
\end{equation}
with $\Sigma^{A/R}$ the advanced/retarded $e$-ph self-energy~\cite{bernardiFirstprinciples2016}. The diagonal components with $n\!=\!n'$ give the renormalized microscopic SRTs, $\tau^{\beta}_{nn\kk} = \tau^\text{e-ph}_{n\kk}\, \Lambda^{\beta}_{nn\kk}(\varepsilon_{n\kk})$ entering Eq.~(\ref{eq:srt}). 
We implement and solve Eqs.~(\ref{eq:bse})-(\ref{eq:dressed_time}) in our {\sc perturbo} code~\cite{zhouPerturbo2021} 
(see Supplemental Material~\cite{supp_mat}).
\\
\indent
The ground state, band structures, and phonon dispersions are obtained using {\sc Quantum ESPRESSO}~\cite{giannozziQUANTUM2009}.
We employ {\sc perturbo}~\cite{zhouPerturbo2021} to compute and interpolate the $e$-ph matrix elements and spin matrices, using a method described in Ref.~\cite{parkSpinphonon2020}, starting from spinor Wannier functions from the {\sc wannier90} code~\cite{pizziWannier902020}. 
We model all materials in the intrinsic (i.e., undoped) limit, and accordingly compare our results with experiments carried out on undoped samples. Additional details are provided in Supplemental Material~\cite{supp_mat}. 
\\
\indent
Using this formalism, in Fig.~\ref{fig:srt_temp} we compute the \textit{macroscopic} SRTs in Eq.~(\ref{eq:srt}) as a function of temperature for Si, GaAs, and monolayer WSe$_2$. In Si, a centrosymmetric material where spin decoherence is governed by the EY mechanism, the results are in excellent agreement with experiments~\cite{appelbaumElectronic2007,lepineSpin1970,lancasterSpinlattice1964} in the 100$-$300~K temperature range.
For example, the SRT computed at 300~K is 6.1~ns, in remarkable agreement with the 6.0~ns value measured in Ref.~\cite{lancasterSpinlattice1964}. Due to the dominant EY mechanism, in this case the sRTA, which neglects spin precession, also gives accurate SRTs.
\\
\indent %
In GaAs, the SOC induces a small ($\sim$1 meV) splitting in the conduction band, so spin relaxation is dominated by the DP mechanism~\cite{mowerDyakonovPerel2011}.
Figure~\ref{fig:srt_temp}(b) shows our calculated SRTs for electrons in GaAs as function of temperature; the excellent agreement with experiments~\cite{oertelHigh2008,kimelRoomtemperature2001,dzhioevSuppression2004,laiTemperature2006} is a strong evidence that the spin-phonon BSE describes correctly the DP mechanism.
By contrast, the sRTA, which captures only the EY mechanism, clearly fails in GaAs, predicting SRTs an order of magnitude greater than experiments.
\\
\indent
Our spin-phonon BSE achieves a similar accuracy for calculations on hole carriers. In
Fig.~\ref{fig:srt_temp}(c), we compute the SRTs for hole spins in monolayer $\text{WSe}_2$, obtaining excellent agreement with all available experimental results between 20$-$90~K~\cite{ersfeldUnveiling2020,liValley2021,yanLong2017,gorycaDetection2019}. 
Note that the valence band of $\text{WSe}_2$ has a large ($\sim$0.4~eV) splitting due to SOC, leading to a precession rate far greater than the hole $e$-ph collision rates; in this strong precession regime, the spin dynamics is controlled by the diagonal part of the spin vertex and the DP mechanism becomes irrelevant, so EY spin decoherence dominates the SRTs.
Conversely, for heavy holes in GaAs (see Fig.~\ref{fig:srt_temp}(d)) both EY and DP spin decoherence are important. 
The agreement with experiment is noteworthy in this regime where both mechanisms are relevant: our computed SRT for holes in GaAs at 300~K is 200~fs, versus a 110~fs value measured by Hilton et al.~\cite{hiltonOptical2002}.
\\
\indent
A key distinction between the EY and DP mechanisms is their dependence on the $e$-ph collision time, $\tau_{n\kk}^\text{e-ph}$ in Eq.~(\ref{eq:srt}): the SRT is proportional to $\tau_{n\kk}^\text{e-ph}$ for EY, and inversely proportional to $\tau_{n\kk}^\text{e-ph}$ for DP. 
Our spin-phonon BSE can capture both of these trends, as we show in Fig.~\ref{fig:srt_temp}(e) by artificially increasing $\tau_{n\kk}^\text{e-ph}$ (%
by multiplying the $e$-ph matrix elements through a constant) and recomputing the SRTs at 300~K for all four cases.
In Si and $\text{WSe}_2$, where EY spin decoherence is dominant, we find that the recomputed SRTs are directly proportional to the $e$-ph collision time, consistent with the EY mechanism~\cite{elliottTheory1954,yafetFactors1963}. 
Conversely, for electron spins in GaAs, the SRTs are nearly inversely proportional to the $e$-ph collision time (see Fig.~\ref{fig:srt_temp}(e)), in agreement with the DP mechanism~\cite{dyakonov1972spin}. (Note that the computed trend slightly deviates from the conventional DP inverse proportionality because EY decoherence, although weak, is still present.)  
For hole spins in GaAs, the recomputed SRTs exhibit a trend intermediate between pure EY and DP, further supporting our conclusion that both mechanisms are important for hole spins in GaAs~\cite{shenHole2010,kraussUltrafast2008}.
\\
\indent
Spin precession in the DP mechanism is induced by the SOC field, which is proportional to the band splitting for each electronic state. To examine the role of DP spin decoherence, we artificially vary the SOC band splitting $\Delta E$ and for each new value we recompute the SRTs (see Fig.~\ref{fig:srt_temp}(f)). 
For $\text{WSe}_2$, varying the SOC band splitting has no effect on the SRTs, showing that spin decoherence is controlled by the EY mechanism. 
For electrons in GaAs, the SRTs are highly sensitive to the SOC splitting, a clear evidence that our formalism can capture the dominant DP mechanism. This dependence is weaker than in the conventional trend for pure DP, $\tau^{(s)}\propto 1/(\Delta E)^2$, due to the coexistence of EY decoherence. 
For hole carriers in GaAs, the SRTs are less sensitive to the SOC splitting than for electrons, as the decoherence originates from a balanced combination of both EY and DP mechanisms.
This analysis also shows that band structure calculations accurately describing the SOC band splitting are essential to predict spin precession and DP decoherence. %
\\
\indent
\begin{figure*}[t!]
\includegraphics[width=1.0\textwidth]{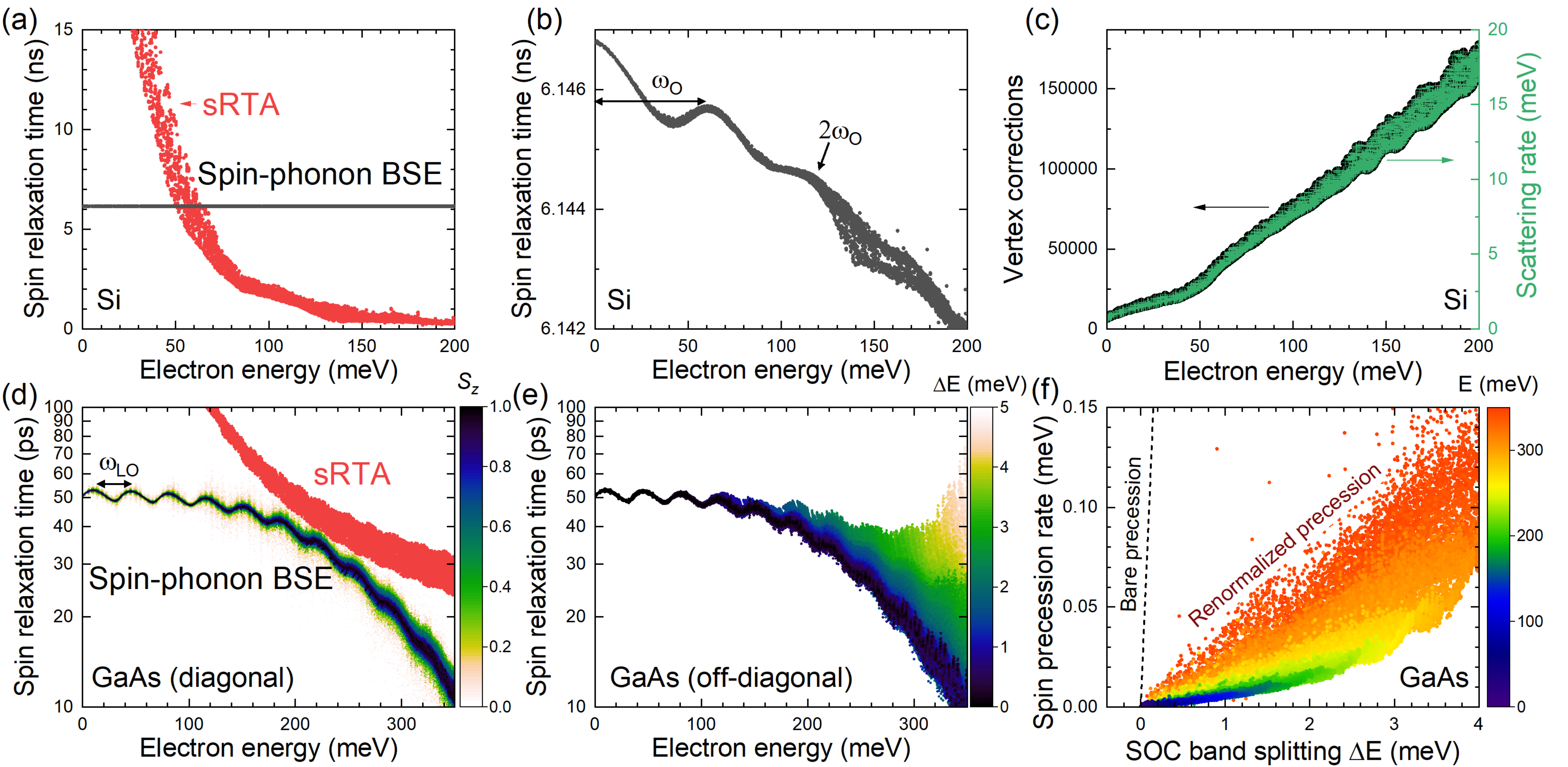} %
\caption{Microscopic spin decoherence.
(a) Microscopic electron SRTs in Si as a function of conduction band energy, computed with the spin-phonon BSE (black) and sRTA (red). 
(b) Zoom-in of the spin-phonon BSE results in (a). 
(c) Vertex corrections $\Lambda_{nn\kk}$ in Si (black dots) compared with the inverse $e$-ph collision times (green crosses). 
(d) Microscopic electron SRTs in GaAs from the spin-phonon BSE, shown as a function of conduction band energy and overlaid with a color map of the expectation value of $S_z$ for each electronic state; the sRTA results (red) are given for comparison. 
(e) Microscopic off-diagonal SRTs, $\tau_{nn'\kk}$ in Eq.~(\ref{eq:dressed_time}), overlaid with a color map of the SOC band splitting $\Delta E$. 
(f) Renormalized electron SPRs in GaAs, $\omega_{nn'\kk}$ in Eq.~(\ref{eq:dressed_time}), plotted as a function of SOC band splitting and overlaid with a color map of the conduction band energy; the bare electron SPRs (black dashed line) are given for comparison. All results are computed at 300~K, and the zero of the energy axis is the conduction band minimum.
}\label{fig:micro}
\end{figure*}
The phonon-induced renormalization greatly modifies the microscopic spin dynamics. 
Figure~\ref{fig:micro}(a) compares sRTA and spin-phonon BSE calculations of the \textit{microscopic} electron SRTs, $\tau^{(s)}_{n\kk} = \tau_{n\kk}^\text{e-ph}\, \Lambda_{nn\kk}^z(\varepsilon_{n\kk})$ defined below Eq.~(\ref{eq:dressed_time}), in Si at 300~K for energies near the conduction band minimum. 
The sRTA results are strongly energy dependent, with an unphysical divergence at low energy.
By contrast, the results from the spin-phonon BSE are nearly energy independent. The vertex correction makes spins with similar energy relax on the same time scale $-$ a constant value of 6.1~ns nearly equal to the macroscopic SRT $-$ and overcomes the limitations of the sRTA. 
A closer examination of the SRTs from the spin-phonon BSE (see Fig.~\ref{fig:micro}(b)) reveals an oscillatory pattern with a period of $\omega_{O}\approx60$~meV, the energy of an optical phonon with strong $e$-ph coupling; this pattern disappears when optical phonons are neglected. 
This oscillation is a manifestation of the self-consistency of the spin-phonon BSE and its ability to capture strong coupling effects beyond lowest-order perturbation theory. 
We observe the same energy dependence and SRT oscillations due to optical phonons for hole spins in $\text{WSe}_2$. 
\\
\indent
Figure~\ref{fig:micro}(c) shows the computed vertex correction $\Lambda^z_{nn\kk}(\varepsilon_{n\kk})$ as a function of energy in Si. %
The vertex correction from the $e$-ph interactions is colossal $-$ relative to the bare spin, it is of order $\Lambda-1 \approx 10^5$, and thus eight orders of magnitude greater than the corresponding vertex correction due to photons in vacuum~\cite{schwingerQuantumElectrodynamics1948,hannekeNew2008} (with value $\Lambda-1 \approx 1.16 \cdot 10^{-3}$, which corrects the electron magnetic moment). %
The energy dependence of the vertex correction is nearly identical to that of the inverse $e$-ph collision times,  
thus explaining the origin of the constant trend with energy of the microscopic SRTs. 
We find large vertex correction values ($10^2-10^5$) also in GaAs and WSe$_2$.
These giant values account for the large differences between $e$-ph collision times (femtoseconds) and SRTs (nanoseconds) in condensed matter, and are key to accurately predicting long spin coherence times of interest in quantum technologies. 
\\
\indent 
In GaAs, due to the Dresselhaus SOC band splitting, the bare spin vertex $s_{nn'\kk}$ acquires large off-diagonal ($n\!\ne\!n'$) components that precess in the effective SOC magnetic field with a \textit{bare} SPR of $\varepsilon_{n\kk} \!-\! \varepsilon_{n'\kk}$. 
While the macroscopic SRTs in Eq.~(\ref{eq:srt}) are determined only by the band diagonal components $s_{nn\kk}$, the spin-phonon BSE couples the diagonal and off-diagonal components via Eq.~(\ref{eq:bse}), so spin precession modifies the SRTs. 
The microscopic SRTs for electrons in GaAs (see Fig.~\ref{fig:micro}(d)) exhibit trends similar to Si $-$ the renormalized SRTs are nearly energy independent near the band edge, in contrast with the rapidly varying SRTs predicted by the sRTA; an oscillating pattern is evident with period equal to the 30~meV longitudinal optical (LO) phonon energy, a signature of strong coupling with LO phonons~\cite{zhouInitio2016}.
Yet, due to the spin precession, we also observe unique trends not found in Si.
The SRTs decrease at higher energies due to the increasing spin precession (the SOC band splitting increases with energy), a manifestation of DP spin decoherence. In addition, the SRTs are strongly state dependent as states with a smaller spin component along the quantization axis, shown with lighter colors in Fig.~\ref{fig:micro}(d), are subject to stronger precession.
\\
\indent
The relaxation of the off-diagonal spin components, quantified by the off-diagonal SRTs $\tau_{nn'\kk}$ in Eq.~(\ref{eq:dressed_time}), reveals additional signatures of the DP mechanism.  
Figure~\ref{fig:micro}(e) shows these off-diagonal electron SRTs for GaAs and highlights their correlation with the SOC band splitting. 
When the band splitting is small (black), precession is negligible and the SRTs are identical to the diagonal SRTs in Fig.~\ref{fig:micro}(d). However, for increasing values of the band splitting (lighter colors), spin precession significantly enhances the SRTs. 
These intriguing microscopic phenomena are encoded in the vertex correction $\Lambda$ in Eq.~(\ref{eq:dressed_time}), which suppresses the real part $1/\tau_{nn'\kk}$ in the denominator, thus slowing down spin relaxation. 
Similarly, the vertex correction significantly slows down spin precession, as shown in Fig.~\ref{fig:micro}(f) for GaAs. Electrons with a bare SPR of 1~meV drop to a $\sim$10$^{-2}$~meV precession rate after renormalization due to phonons. These renormalized SPRs are strongly energy dependent, with higher electron energies leading to faster precession for spins with the same bare SPRs. This microscopic dynamics reveals the rich interplay between spin relaxation and precession in materials. %
\\
\indent
In conclusion, our findings highlight the dramatic effects of phonon-induced renormalization on electron spins in solids. 
Our spin-phonon BSE can capture renormalized spin dynamics beyond relaxation, shedding light on the interplay between the EY and DP spin decoherence mechanisms, and describing their diverse physics on the same footing.
This formalism reveals that the long intrinsic spin coherence times in condensed matter are due to the colossal vertex correction from $e$-ph interactions. %
Our computationally affordable method enables precise predictions of spin decoherence, with broad implications for spin-based quantum technologies and for advancing microscopic understanding of spin dynamics in solids.\\ %

\begin{acknowledgments}
\noindent
This work was supported by the National Science Foundation under Grants No. DMR-1750613 and QII-TAQS 1936350, which provided for method development, and Grant No. OAC-2209262, which provided for code development.
J.P. acknowledges support by the Korea Foundation for Advanced Studies. %
This research used resources of the National Energy Research Scientific Computing Center (NERSC), a U.S. Department of Energy Office of Science User Facility located at Lawrence Berkeley National Laboratory, operated under Contract No. DE-AC02-05CH11231. 
\end{acknowledgments}

\vspace{-5pt}
\bibliographystyle{apsrev4-2}

\end{document}